\documentclass[12pt]{article}

\def\be{\begin{equation}}
\def\ee{\end{equation}}
\def\bea{\begin{eqnarray}}
\def\eea{\end{eqnarray}}
\def\bean{\begin{eqnarray*}}
\def\eean{\end{eqnarray*}}
\def\bt{\begin{table}[ht]}
\def\et{\end{table}}
\def\ba{\begin{array}}
\def\ea{\end{array}}
\def\bc{\begin{center}}
\def\ec{\end{center}}
\def\nue{\nu _e}
\def\numu{\nu _{\mu}}
\def\nutau{\nu _{\tau}}

\def\jrn#1#2#3#4{{#1} {\bf #2}, #3 (#4)}
\def\NPB{Nucl. Phys. B}
\def\PLB{Phys. Lett. B}
\def\PR{Phys. Rept.}
\def\PRD{Phys. Rev. D}
\def\PRL{Phys. Rev. Lett.}
\def\PTP{Prog. Theor. Phys.}

\begin{document}


\pagestyle{empty}

\vspace*{1.5cm}
\begin{center}
\LARGE{Family Hierarchy and Large Neutrino Mixings\\[20mm]}
\large{Fu-Sin~Ling\footnote{e-mail: fsling@phys.ufl.edu} and
Pierre Ramond\footnote{e-mail: ramond@phys.ufl.edu}\\[8mm]}
\it{Institute for Fundamental Theory\\
Department of Physics, University of Florida,\\ 
Gainesville, FL, 32611, USA\\[15mm]}
\large{\rm{Abstract}} 
\\[7mm]
\end{center}

\begin{center}
\begin{minipage}[h]{14cm}
The recent neutrino data seem to favor two large and one small mixing 
angles and a hierarchy of their squared mass differences.
We discuss these within the context of hierarchical neutrino masses.
We show that this scheme suggests a specific neutrino mass matrix
with mild fine-tuning. We then present a Froggatt-Nielsen model
that reproduces this matrix. 
\end{minipage}
\end{center}
\newpage

\pagestyle{plain}


\section{Introduction}

Recently, numerous neutrino experiments have confirmed the first
true signal of physics beyond the standard model.
The flavor conversion of neutrinos, or neutrino oscillations has been 
observed in both the atmospheric and the solar neutrinos. Therefore, unlike in the Standard Model, neutrinos
do have mass and mixings.

The atmospheric neutrino data are likely explained
by a $\numu - \nutau$ oscillation with 
parameters~\cite{toshito, SKatm}
\bean
\Delta m^2 &\simeq& 2.5 \cdot 10^{-3} \; eV^2 \\
\sin^2 2 \theta &\geq& 0.8
\eean

For the solar neutrinos deficit, there are four different solutions,
called Large Mixing Angle (LMA), Small Mixing Angle (SMA), "Low", 
and the vacuum oscillation solution (VO). Typical values of the
mixing angle and the $\Delta m^2$ at best fit are shown in 
table~\ref{solnutable}~\cite{solnusol}
\bt
\bc
\begin{tabular}{|c|c|c|}
\hline
Solution & $\tan^2 \theta$ & $\Delta m^2 \; (eV^2)$\\
\hline
LMA & $3 \cdot 10^{-1}$ & $3 \cdot 10^{-5}$ \\
SMA & $10^{-3}$ & $7 \cdot 10^{-6}$ \\
LOW & $6 \cdot 10^{-1}$ & $10^{-7}$ \\
VO & $3 \cdot 10^{-1}$ or $3$ & $8 \cdot 10^{-11}$ \\
\hline
\end{tabular}
\ec
\caption{\label{solnutable} Typical values of $\tan ^2 \theta$ and 
$\Delta m^2$ for the various solar neutrinos solutions at best fit.}
\et
These solutions are obtained by combining the rates observed
at different types of experiments (Homestake, Gallex, Sage, 
SuperKamiokande (SK), SNO). However, in addition, each solution
 has a characteristic signature in various observables
like the energy spectrum, the day-night asymmetry, the seasonal
variation, or the zenith angle dependence. 
Even though a possible distortion is small and difficult to measure,
the huge statistics accumulated by SK made it possible to almost
exclude the VO and the SMA solutions~\cite{SKsol}. 
Moreover, the first results of the ongoing experiment SNO not 
only demonstrate the existence of flavor conversion of the solar
neutrinos independently of any solar model~\cite{SNO1}, 
but also seem to favor the LMA solution as some
day-night effect is measured~\cite{SNO2}. The third signal of neutrino oscillations from
LSND~\cite{LSND} needs confirmation.

If we are to believe these data, and if we assume a framework
with only three flavors ($\nue$, $\numu$, $\nutau$), then the
neutrino sector exhibits two large mixing angles, and one small
mixing angle, imposed by the CHOOZ~\cite{CHOOZ} experiment.
At first sight, mass hierarchies and mixing patterns are
completely different in the lepton and in the quark sectors.
For the quarks, the observed mixing matrix is given by the
CKM matrix, which is almost the unit matrix. 
The order of magnitude of the off-diagonal elements are 
given as a power of the Cabibbo angle $\lambda _c$ in the 
Wolfenstein parametrization. 
The corresponding MNS matrix in the lepton sector, which
is now measured in the neutrino oscillation experiments,
does not however seem to fit in such parametrization scheme
if two large mixing angles are indeed observed.

In this paper, we analyze the requirements that are needed
in the effective neutrino Majorana mass matrix in order to give
rise to two large and one small mixing angles, in the context of 
hierarchical neutrino masses. 
The derived structure enables us 
to construct a model of family hierarchy based on additional
Abelian family symmetries, in the spirit of~\cite{FM01,FM02},
which is compatible with the neutrino observations.
It turns out that the hierarchies and mixings among leptons and 
quarks are highly interrelated in this model, in sharp contrast
with the picture that appears at first glance.

The paper is organized as follows. In section 2, we analyze
the effective $3 \times 3$ neutrino mass matrix. In section 3,
we present a new model of family symmetry and its predictions
for  the neutrinos, charged leptons and  quarks masses and mixings.
The internal consistency of the model, including anomaly cancellation is then briefly discussed.

\section{Neutrino mass hierarchies and large mixing angles}

In this section, we derive the structure of the $3 \times 3$ 
effective neutrino mass matrix assuming a hierarchical pattern
of the neutrino masses.

As neutrino experiments are only sensitive
to the values of $\Delta m^2$, but not to the absolute
values of the neutrinos masses, the range of the neutrino masses 
could in principle be different from the measured splittings.
For example, if $m_{\nu} \sim 1 \; eV$, which could be useful
for cosmology purposes, the observed $\Delta m^2$
would result in a hyperfine-like splitting
structure in the neutrino sector, totally different from what
is observed in the other fermionic sectors.

In what follows, we assume that a mass hierarchy takes place
in the neutrino sector, and that it is responsible of the observed
hierarchy between $\Delta m^2_\odot$ for solar neutrinos 
and $\Delta m^2_\oplus$ for atmospheric 
neutrinos~\footnote{Therefore, we ignore as well the possibility
of the so-called inverted hierarchy, where 
$m_1 \simeq -m_2 \gg m_3$, and $\Delta m^2_\odot = m_1^2 - m^2_2$,
which also gives a large atmospheric mixing 
angle~\cite{king}}.
For example, for the LMA solution, we have
\be
\frac{\Delta m^2_\odot}{\Delta m^2_\oplus} \simeq 10^{-2}
\ee 

Therefore, we suppose that the neutrinos masses can be written as
$\widetilde{M}_{\nu} = m_0 \cdot 
diag(\beta \lambda ^b ,\alpha \lambda ^a, 1)$,
where $\lambda$ is some small parameter, 
$b \geq a >0$ are integers
and $\alpha$, $\beta$ are numerical coefficients of order 1.
Then, in the charged leptons mass eigenstates basis,
the full effective mass matrix will be
\be
M_{\nu} = U_{MNS} \cdot \widetilde{M}_{\nu} \cdot U_{MNS}^t
\ee
with
\be
U_{MNS} \simeq \left( 
\ba{ccc}
\cos \theta _\odot & -\sin \theta _\odot & \ll 1 \\
\sin \theta _\odot \cos \theta _\oplus & 
\cos \theta _\odot \cos \theta _\oplus & -\sin \theta _\oplus \\
\sin \theta _\odot \sin \theta _\oplus & 
\cos \theta _\odot \sin \theta _\oplus & \cos \theta _\oplus \\
\ea
\right)
\ee
The structure found for $M_{\nu}$ is 
\be
M_{\nu} \simeq m_0 \cdot \left(
\ba{ccc}
\gamma \lambda ^c & \delta _1 \lambda ^d & \delta _2 \lambda ^d \\
\delta _1 \lambda ^d & \sin^2 \theta _\oplus & 
-\sin \theta _\oplus \cos \theta _\oplus \\
\delta _2 \lambda ^d & -\sin \theta _\oplus \cos \theta _\oplus &
\cos^2 \theta _\oplus \\
\ea
\right)
\label{matrix}
\ee
where $\gamma$, $\delta _1$ and $\delta _2$ are some numerical
factors. We see that the large value of the atmospheric mixing angle
forces all the elements in the 2-3 sub-block of $M_\nu$ to be 
${\cal O}(1)$. Moreover, the hierarchy of the neutrinos masses
forces the determinant of the 2-3 sub-block, referred below
as the sub-determinant $D_1$ to be zero, {\it i.e.} $D_1 \ll 1$.
The effect of the second large mixing angle is more subtle to
analyze. Basically, it amounts to a relation between $D_1$
and the full $3 \times 3$ determinant of coefficients $D_0$. 

A structure of the neutrino mass matrix similar to the one found
in~(\ref{matrix}) can be derived in models with an Abelian family symmetry. 
However, the additional condition on the sub-determinant $D_1$
is not generic in models of family hierarchy. 
Different possible mechanisms to solve this problem have been
advocated~\cite{hierLMA}. In what follows, we also focus on the
conditions on $D_0$ and $D_1$ to get a large solar mixing angle
in addition to the large atmospheric angle. 

As an illustrative example, we take the prediction for the
neutrino mass matrix in the model of ref.~\cite{FM01}, 
\be
M_\nu^{(1)} \sim m_0 \left(
\ba{ccc}
\lambda^6 & \lambda^3 & \lambda^3 \\
\lambda^3 & 1 & 1 \\
\lambda^3 & 1 & 1 \\
\ea
\right)
\ee 
where $\lambda$ is of the order of the Cabibbo angle 
$\lambda \simeq \lambda _c$. 
Generically, {\it i.e.} without any assumption on the numerical
prefactors, the solar mixing angle obtained is small, 
$\theta _\odot \sim {\cal O}(\lambda^3)$, and the mass spectrum is 
$m_\nu \sim m_0 \cdot (\lambda^6,1,1)$, which does not result in the 
desired hierarchy of the $\Delta m^2$. 
As noticed in ref.~\cite{FM03}, two large mixing angles and
the hierarchy of $\Delta m^2$ can be obtained,  but it requires 
to fine-tune the sub-determinant to order ${\cal O}(\lambda^3)$. 
More precisely, we can
summarize (see table~\ref{finetune}) the mixing and hierarchies 
obtained when $D_0 \sim {\cal O}(1)$
and $D_1 \sim {\cal O}(\lambda^p)$, with $p \geq 0$ integer.
\bt
\bc
\begin{tabular}{|c|c|c|c|}
\hline
$D_0 \sim {\cal O}(1)$, $D_1 \sim {\cal O}(\lambda^p)$ & 
$m_\nu$ ($\cdot m_0$) & $\theta _\odot$ & 
 $\Delta m^2_\odot / \Delta m^2_\oplus$ \\
\hline
$p=0$ & $\lambda^6,1,1$ & $\lambda^3$ & ${\cal O}(1)$ \\
$1 \leq p < 3$ & $\lambda^{6-p}, \lambda^p, 1$ & 
$\lambda^{3-p}$ & $\lambda^{2p}$ \\
$p=3$ & $\lambda^3, \lambda^3, 1$ & 
${\cal O}(1)$ & $\lambda^6$ \\
$3<p<6$ & $\lambda^3, \lambda^3, 1$ & 
$\pi/4$ & $\lambda^{p+3}$ \\
$p \geq 6$ & $\lambda^3, \lambda^3, 1$ & 
$\pi/4$ & $\lambda^9$ \\
\hline
\end{tabular}
\ec
\caption{\label{finetune} Neutrino mixing and hierarchies
from $M_\nu^{(1)}$ as a function of $p$.}
\et

As we can see, with the mass matrix $M_\nu^{(1)}$, two large 
mixing angles result only if $D_1$ is precisely fine-tuned to
order ${\cal O}(\lambda^3)$. For $p>3$, the solar angle 
$\theta _\odot$ becomes maximal, and this is excluded 
at 99\% C.L. by the experimental data~\cite{solnusol}.
Moreover, for $p=3$, the value of the ratio 
$\Delta m^2_\odot / \Delta m^2_\oplus$  
would favors the ``LOW" solution which only exists at 99\% C.L. 
if all observed rates are combined and day-night asymmetry 
is taken into account.
Therefore, we conclude that two large mixing angles cannot be reproduced
by the matrix $M_\nu^{(1)}$ without strong fine-tuning. 

On the other hand, the following neutrino mass matrix, $M_\nu^{(2)}$, 
which has the same structure as $M_\nu^{(1)}$, 
is in good agreement with all the neutrino data without strong fine-tuning.
\be
M_\nu^{(2)} \sim m_0 \left(
\ba{ccc}
\lambda^2 & \lambda & \lambda \\
\lambda & 1 & 1 \\
\lambda & 1 & 1 \\
\ea
\right)
\ee
with again $\lambda \simeq \lambda _c$.
The CHOOZ constraint is easily satisfied because 
$U_{e3} \sim {\cal O}(\lambda)$~\footnote{In models which generically predict
three large mixing angles, like models of neutrino anarchy~\cite{berger},
the ChOOZ constraint becomes critical}.
When $D_0 \sim {\cal O}(1)$ and $D_1 \sim {\cal O}(\lambda)$ 
(which is a mild assumption),
one gets two large mixing angles and 
$\Delta m^2_\odot / \Delta m^2_\oplus \sim {\cal O}(\lambda^2)$. 
The properties of the mass matrix $M_\nu^{(2)}$ have also been investigated 
using a random coefficients generator~\cite{sato, vissani}
, and it appears that the probability
of a successful phenomenological description of the neutrino data becomes
maximal for $\lambda \simeq \lambda _c$~\cite{vissani}.
The same structure for the neutrino mass matrix also appears
in several different contexts~\cite{carone,huber,maekawa}.

Moreover, with the matrix $M_{\nu}^{(2)}$, there is a special 
case which gives rise to two large mixing 
angles, which may become relevant as the neutrino masses are generated
through the seesaw~\cite{SEESAW} mechanism. 
We note from table~\ref{finetune} that a large solar mixing angle
arises when the two light eigenvalues $\mu _1$ and $\mu _2$
are of the same order of magnitude.
\be
\mu _1 \sim \mu _2 \; \Longrightarrow \; 
\sin\,\theta _\odot \sim {\cal O}(1) 
\ee
The characteristic equation for $M_\nu^{(2)}$ is of the form
\be
\mu^3 + {\cal O}(1) \cdot \mu^2 + 
\left( D_1 + {\cal O}(\lambda^2) \right)\mu + D_0 \cdot \lambda^2 = 0
\ee
Therefore, the solar mixing angle becomes large 
when $D_0 \sim {\cal O}(\lambda^2)$ and 
$D_1 \sim {\cal O}(\lambda^2)$ because we have 
$\mu _1 \sim \mu _2 \sim {\cal O}(\lambda^2)$ in this case.
The analytic values of the eigenvalues and the mixing matrix
can be found in the appendix.
The reason why this special case might be relevant is simple.
The light neutrino masses are given by a seesaw mechanism.
In the chiral Froggatt-Nielsen type models, the charges of the right-handed
neutrinos are canceled in the seesaw, and the Cabibbo structure of the
effective neutrino mass matrix only depends on the charges of the
lepton doublets $L_i$ (see~\cite{FM01}). 
However, the numerical coefficients, or pre-factors are given by the 
usual see-saw formula
\be
C_\nu = -C_D \cdot C_0^{-1} \cdot C_D^t
\ee
where $C_\nu$, $C_D$ and $C_0$ are the $3 \times 3$ matrices of
pre-factors corresponding to the effective light neutrino mass matrix,
the Dirac mass matrix and the heavy neutrino Majorana mass matrix
respectively. 
Let's now suppose that the coefficients of the second and of the third 
lines of $C_D$, are approximately (to order ${\cal O}(\lambda))$ equal,
so that the matrix $C_D$ can be written as
\be
C_D = \left(
\ba{ccc}
a & b & c \\
d_0 + d_1 \lambda & e_0 + e_1 \lambda & f_0 + f_1 \lambda \\
d_0 - d_1 \lambda & e_0 - e_1 \lambda & f_0 - f_1 \lambda \\
\ea
\right)
\ee
where all the coefficients are ${\cal O}(1)$.
Then, after see-saw, the matrix $C_\nu$ indeed verifies 
$D_0 \sim {\cal O}(\lambda^2)$ and $D_1 \sim {\cal O}(\lambda^2)$.
This form for the matrix $C_D$ can appear if for example 
the Lagrangian is symmetric under 
the exchange of the lepton doublets of the second and the third family,
so that we call this symmetry $P_{23}$.
If instead the lagrangian is antisymmetric, the corresponding approximate 
$-P_{23}$ symmetry would have the same consequences. 
On the other hand, because $\lambda$ is not a very small expansion parameter,
the appearance of an approximate $P_{23}$ symmetry in the matrix $C_D$
does not require a fundamental symmetry of the lagrangian. 
We will further address this question in the context
of our family symmetry model.  
If this approximate $P_{23}$ (or $-P_{23}$) discrete symmetry is
realized, then the pattern of eigenvalues becomes
\be
m_0 \cdot (\lambda^2, \lambda^2, 1)
\ee
so that $\Delta m^2_\odot / \Delta m^2_\oplus \sim {\cal O}(\lambda^4)$. 
Moreover, the atmospheric mixing angle is automatically
maximal $\sin^2 2\theta _\oplus \simeq 1 $, 
as suggested by the experimental data. 
We emphasize that a maximal atmospheric mixing angle is in general
not a feature of models with a family symmetry. 
Statistically~\cite{vissani}, the spread is rather large, so that
the experimental bound 
\be
\sin^2 2 \theta _\oplus \geq 0.8
\ee
is in general not automatically satisfied, even if the model predicts
a large atmospheric mixing angle.

To conclude this section, we have found that the matrix $M_{\nu}^{(2)}$
with $\lambda \simeq \lambda _c$ can accommodate in different ways  
all the present data on neutrino oscillations (exception
made for LSND) with very little fine-tuning.
In a context of models with family hierarchies, this structure is 
suggested by the neutrino oscillation data if the LMA solution
to the solar neutrino problem is favored, and the mild measured
hierarchy between the $\Delta m^2$ is reflected in the mild
hierarchical pattern of $M_\nu^{(2)}$. In the rest of the paper, we present a 
Froggatt-Nielsen type model which reproduces our favored  neutrino  mass matrix $(6)$.

\section{A model of family hierarchy}

This model of family hierarchy uses extra Abelian flavor symmetries
to reproduce the observed hierarchies among quarks and leptons, 
in the spirit of an effective
theory, as suggested by Froggatt and Nielsen~\cite{FN}.
It is simpler but analogous in its setup to the model in~\cite{FM01, FM02}.
We refer the interested reader to these papers for the general 
features of the framework, and turn now to the specific points of the
model.

\subsection{The family symmetries}

The gauge sector contains the Standard Model groups 
and two additional Abelian symmetries, $U(1)_{Y_F}$ and $U(1)_X$.
It may also contain some hidden gauge group, 
so that the complete gauge structure looks like
$$
G_{visible} \times U(1)_{Y_F} \times U(1)_X \times G_{hidden}
$$
and
$$
G_{visible} \supseteq SU(3) \times SU(2) \times U(1)_Y
$$

The chiral matter content is given by the representation 
{\bf 16} of $SO(10)$. $SO(10)$ contains one $U(1)$ symmetry out of the 
Standard Model. 
$$
SO(10) \supset SU(5) \times U(1)_V
$$
The {\bf 16} multiplet includes all the fermion fields of the 
Standard Model, plus a right-handed neutrino which is 
necessary to implement the see-saw mechanism.
\be
{\bf 16} = {\bf 10}_1 + {\bf \bar{5}}_{-3} + {\bf 1}_5
\ee
The symmetry $U(1)_{Y_F}$ is traceless over the three families of
quarks and leptons and is non-anomalous. Its form is given by
\be
Y_F=\eta _1(1,0,-1)+\eta _2(0,1,-1)
\ee
with
\be
\eta _1 = -\frac{V+7V'}{6}\ ,\qquad
\eta _2 = -\frac{V+V'}{6}
\ee
and $V'=1$ for all the matter fields in the {\bf 16} multiplet
($V'$ may be viewed as the extra $U(1)$ in the embedding 
$E_6 \supset SO(10) \times U(1)_{V'}$). One can notice
that the family symmetry $U(1)_{Y_F}$ lies outside of $SU(5)$.
As a result, all the different fields in the same $SU(5)$ multiplet
have the same family structure. Therefore, the hierarchies and mixings
predicted in the lepton and in the quark sectors are strongly 
related in this model.

The symmetry $U(1)_X$ is family independent and necessarily
anomalous. Its presence is needed in order to construct the 
right-handed neutrinos Majorana mass matrix. It also participates
in the intrafamily hierarchy, namely in the ratio $m_b/m_t$ of the
bottom quark mass to the top quark mass. 
Over the three chiral families, it has the following form
\be
X = \alpha + \beta V
\ee
where $\alpha$ and $\beta$ are coefficients to be determined.
The anomalies linear in $X$ in the effective theory are taken care of 
by the Green-Schwarz anomaly cancellation mechanism~\cite{GSM}. 
All the other anomalies in the model must vanish.

To construct the quarks and leptons mass matrices,
the model also contains two pairs pairs $(H_u,H'_d)$ and $(H'_u,H_d)$
in the Higgs sector. For the sake of anomaly cancellation, 
they are vector-like with respect to the additional Abelian symmetries.
Among them, only $H_u$ and $H_d$ will be coupled to the matter fields
to form the supersymmetric and gauge invariants in the superpotential.

Two chiral superfields $\theta _1$ and $\theta _2$ are needed to mediate
the breaking of the flavor symmetries. We express their charges in the 
matrix $A$ 
\be
A = \left(
\ba{ccc}
X(\theta _1) & X(\theta _2) \\
Y_F(\theta _1) & Y_F(\theta _2) \\
\ea
\right)
\ee
These charges are constrained by the vacuum structure, 
and the phenomenological requirement that the coupling of the fields 
$\theta$ to the Standard Model invariants reproduce the observed 
hierarchies. These constraints are sufficient to fix the matrix $A$
since all the elements of $A$ and $A^{-1}$ must be integer. We have
\be
A = \left(
\ba{cc}
1 & 0 \\
-1 & 1 \\
\ea
\right)
\hspace{.5cm}
{\rm and}
\hspace{.5cm}
A^{-1} = \left(
\ba{cc}
1 & 0 \\
1 & 1 \\
\ea
\right)
\ee

\subsection{Predictions of the model}

We now successively give the predictions of the model for the
up quarks, down quarks, charged leptons and neutrinos.

\subsubsection{Up-quark masses}

The up-quark  masses correspond to the invariant $S = Q \bar{u} H_u$.
The charge $Y_F (H_u)$ is fixed by the fact that the top quark mass
appears unsuppressed in the superpotential,
\be
Y_F (H_u) = -\frac{10}{3}
\hspace{.5cm}
{\rm and}
\hspace{0.5cm}
Y_F (H'_d) = \frac{10}{3}
\ee
We also have
\be
X^{[u]} \equiv X(Q \bar{u} H_u) = 0
\ee
Therefore, the Yukawa matrix for the up quarks has 
the following structure
\be
Y^{[u]} \sim \left(
\ba{ccc}
\lambda^6 & \lambda^5 & \lambda^3 \\
\lambda^5 & \lambda^4 & \lambda^2 \\
\lambda^3 & \lambda^2 & 1 \\
\ea
\right)
\ee
leading to the ratios
\be
\frac{m_u}{m_t} \sim \lambda^6
\hspace{1cm}
\frac{m_c}{m_t} \sim \lambda^4
\ee
which are in agreement with phenomenology~\footnote{Other
models~\cite{FM01, berger} 
predict a smaller ratio $\frac{m_u}{m_t} \sim \lambda^8$,
but both values are consistent with phenomenology}.

\subsubsection{Down-quark masses}

Holomorphy of the terms containing the invariant $S=Q \bar{d} H_d$
forces the charge $Y_F (H_d)=n$ to be an integer, so that $H_d$
cannot be the vector partner of $H_u$. 
If $X^{[d]} \equiv X(Q \bar{d} H_d)$, the Yukawa matrix
for the down quarks has the following structure in the absence of 
supersymmetric zeros ({\it i.e.} $X^{[d]} \leq 0$ is an integer 
and $X^{[d]}+n+2 \leq 0$)
\be
Y^{[d]} \sim \lambda^{-2 X^{[d]}-2-n} \left(
\ba{ccc}
\lambda^4 & \lambda^3 & \lambda^3 \\
\lambda^3 & \lambda^2 & \lambda^2 \\
\lambda & 1 & 1 \\
\ea
\right)
\ee
leading to the interfamily hierarchy
\be
\frac{m_d}{m_b} \sim \lambda^4
\hspace{1cm}
\frac{m_s}{m_b} \sim \lambda^2
\ee
Diagonalization of the Yukawa matrices $Y_u$ and $Y_d$ yields
the correct Cabibbo structure for the CKM mixing matrix.
\be
U_{CKM} \sim \left(
\ba{ccc}
1 & \lambda & \lambda^3 \\
\lambda & 1 & \lambda^2 \\
\lambda^3 & \lambda^2 & 1 \\
\ea
\right)
\ee
In this model, the intrafamily hierarchy between $m_t$ and $m_b$
is not predicted. We have
\be
\frac{m_b}{m_t} \sim \cot \beta \lambda^{-2X^{[d]}-2-n}
\ee
which has to be matched with the observed value
\be
\frac{m_b}{m_t} \sim \lambda^3
\ee

\subsubsection{Charged lepton masses}

As the family charges of this model are unified into $SU(5)$
multiplets, it follows that the charge structure of the invariants
$L \bar{e} H_d$ and $Q \bar{d} H_d$ are the same. Therefore,
\be
Y^{[e]} \sim \lambda^{-2 X^{[d]}-2-n} \left(
\ba{ccc}
\lambda^4 & \lambda^3 & \lambda^1 \\
\lambda^3 & \lambda^2 & 1 \\
\lambda^3 & \lambda^2 & 1 \\
\ea
\right)
\ee  
leading to the ratios
\be
\frac{m_e}{m_\tau} \sim \lambda^4
\hspace{1cm}
\frac{m_\mu}{m_\tau} \sim \lambda^2
\ee
A smaller ratio $\frac{m_e}{m_\tau} \sim \lambda^{5-6}$ would 
somehow be in better agreement with the measured masses, 
but $\lambda^4$ is not incompatible~\footnote{The main problem
with the $SU(5)$ unification of the charges is that the observed
$m_e/m_\mu$ and $m_d/m_s$ ratios are different by a factor 10.
However, taking the ratio of the two light masses might not be a good
"measure" of the agreement between the model and the data, because
many order one coefficients are involved in this ratio.
Rather, for $m_d \simeq 6 \, MeV$, the ratios $m_e/m_\tau$
and $m_d/m_b$ are separated by a factor 4.5 only.

A mechanism to deal with this problem is also proposed in 
ref.~\cite{ellis}}.
For the intrafamily hierarchy, the model predicts
\be
\frac{m_b}{m_\tau} \sim 1
\ee
which indeed corresponds to what is observed. 
We take this fact as an experimental hint that the family symmetries 
are beyond SU(5).

\subsubsection{Neutrino masses}

The light neutrinos masses are obtained after seesaw. 
The Majorana mass term for the right-handed neutrinos is based
on the invariant $M \bar{N} \bar{N}$. Similarly to the model of
ref.~\cite{FM02}, the flat direction analysis of the 
vacuum (see~\cite{flat}) fixes the $X$-charge of $\bar{N}$
\be
X(\bar{N})=-\frac{3}{2}
\ee 
The right-handed neutrino Majorana mass matrix then contains a
harmless supersymmetric zero in position $33$, and is given by
\be
M_0 \sim M \left(
\ba{ccc}
\lambda^{10} & \lambda^9 & \lambda^5 \\
\lambda^9 & \lambda^8 & \lambda^4 \\
\lambda^5 & \lambda^4 & 0 \\
\ea
\right)
\ee
The order of magnitude of the massive right-handed neutrinos is 
therefore given by
\be
m_{\bar{N}_1} \sim M \lambda^{10}
\hspace{1cm}
m_{\bar{N}_2} \sim M \lambda^8
\hspace{1cm}
m_{\bar{N}_1} \sim M
\ee
It is interesting to notice that the mass of the heaviest 
right-handed neutrino is not suppressed by any power of $\lambda$.

The right-handed neutrinos are coupled to the left-handed neutrino
through the invariant $L \bar{N} H_u$. We notice that, with 
$X = \alpha + \beta V$, we have
\be
X^{[\nu]} \equiv X(L \bar{N} H_u) = 
X(L)+X(\bar{N})-X(Q)-X(\bar{u}) = 0
\ee
for any values of $\alpha$ and $\beta$. As a result, the Dirac mass
matrix is given by ($<H_u^0> = v_u$)
\be
m_D \sim v_u \left(
\ba{ccc}
\lambda^6 & \lambda^5 & \lambda \\
\lambda^5 & \lambda^4 & 1 \\
\lambda^5 & \lambda^4 & 1 \\
\ea
\right)
\ee
The light neutrinos mass matrix obtained after see-saw is
\be
m_\nu \sim \frac{v_u^2}{M} \left(
\ba{ccc}
\lambda^2 & \lambda & \lambda \\
\lambda & 1 & 1 \\
\lambda & 1 & 1 \\
\ea
\right)
\ee
The measured value of $\Delta m^2_\oplus$ gives a phenomenological
constraint on the scale $M$. We find
\be
M \sim 10^{16} \; GeV
\ee
which agrees very well with the string scale. 
The typical mass of the lightest right-handed neutrino is then 
$m_{\bar{N}_1} \sim 10^{9-10} \; GeV$.
For the purpose of leptogenesis, this value is somehow too small
in the classical scheme of the decay of right-handed 
neutrinos~\cite{leptogen1}.
But if leptogenesis occurs in a scheme where the Majorana neutrinos are 
only virtual, then the scale found here agrees with the thermodynamical
conditions of the scenario~\cite{leptogen2}.

We now turn back to the question of the significance of an approximate
$P_{23}$ (or $-P_{23}$) symmetry. If this symmetry is somehow 
theoretically motivated, it will appear at the level of the superpotential,
which will be symmetric (or antisymmetric) for the exchange of the 
lepton doublets $L_2$ and $L_3$. This will affect both the Dirac mass term
and the charged leptons mass term. As a result, the muon mass will 
be suppressed by an extra power of $\lambda$, so that
\be
\frac{m_\mu}{m_\tau} \sim \lambda^3
\ee
Moreover, the diagonalization $U_{-1}^\dagger Y^{[e]} V_{-1}$
and $U_0^t m_\nu U_0$ of the charged lepton and light neutrino masses
$Y^{[e]}$ and $m_\nu$ yields a maximal mixing angle $\theta _{23}$ 
in both $U_{-1}$ and $U_0$. These cancel out in the MNS matrix
$U_{MNS} = U_{-1}^\dagger U_0$, and the atmospheric mixing angle
becomes small. Therefore, we conclude that the approximate
$P_{23}$ symmetry is not phenomenologically acceptable in this model,
if it is taken as a fundamental symmetry in the effective theory. 
However, an accidental approximate symmetry can still be viable.

\subsection{Anomaly cancellation}

The consistency of the model depends on anomaly cancellations. There are different possible anomalies that we will analyze in the 
following sequence: Standard Model gauge groups $G_{SM}$ anomalies
($G_{SM}$ is $SU(3)$, $SU(2)$ or $U(1)_Y$), 
mixed anomalies between $G_{SM}$ and $U(1)_{Y_F}$, $U(1)_{Y_F}^3$ anomaly,
anomalies involving $U(1)_X$ and gravitational anomalies.

The anomaly coefficients involving only the Standard Model gauge group vanish
over the chiral matter as well as over the vector-like pairs of Higgs, so
$(G_{SM}G_{SM}G_{SM})=0$.

The anomaly coefficients of the type $(Y_F G_{SM} G_{SM})$ vanish over the
three fermion families because $Y_F$ is traceless. They also vanish on the
Higgs as these are vector-like with respect to $Y_F$.
The anomaly coefficients $(Y_F Y_F G_{SM})$ vanish automatically if $G_{SM}$
is either $SU(3)$ or $SU(2)$. 
The cancellation of the coefficients $(Y_F Y_F Y)$ is ensured over an $SU(5)$
multiplet because $Y_F$ lies outside $SU(5)$. Therefore,
\be
(Y_F Y_F Y)_{{\bf 10}} = (Y_F Y_F Y)_{{\bf \bar{5}}} = 0
\ee
Finally, the right-handed neutrino has zero hypercharge and the vector-like 
Higgs do not contribute. It is worth emphasizing that all the mixed anomaly 
coefficients involving $G_{SM}$ and $Y_F$ vanish over the representation
{\bf 16} of $SO(10)$. Therefore, it is not necessary to extent the
matter content of the model to the representation {\bf 27} of $E_6$ like
in the model of ref.~\cite{FM02}.

The anomaly coefficient $(Y_F Y_F Y_F)$ does not vanish over the three 
chiral families. We find
\be
(Y_F Y_F Y_F)_{chiral} = 39+ \frac{1}{9}
\ee
We also expect some contribution from the $\theta$ sector. These anomalies
need to be cancelled by some additional fields in the model that are singlet
under the Standard Model gauge groups, but carry $X$ and $Y_F$ charges.
The detailed structure of the additional matter fields necessary
to cancel this anomaly is beyond the scope of the present paper, as 
this matter content does not have a phenomenological impact.

As mentioned, the family-independent symmetry $U(1)_X$ is necessarily 
anomalous. The anomalies linear in $X$ are compensated through
the Green-Schwarz anomaly cancellation mechanism~\cite{GSM}.
As a consequence, the value of the Weinberg angle at the 
cut-off $M$ is related to anomaly coefficients~\cite{ibanez}, 
\be
\sin ^2 \theta_W = \frac{C_2}{C_1 + C_2}
\ee
In our model, we have
\bea
C_1 = 10 \alpha \\
C_2 = C_3 = 6 \alpha
\eea
so that the calculated value of $\sin ^2 \theta_W$ is equal to the
canonical value in $SU(5)$ theories,
\be
\sin ^2 \theta_W = \frac{3}{8}
\ee
which is a consequence of the fact that $U(1)_X$ lies outside $SU(5)$.
The anomalies $(X Y_F Y_F)$ and $(XXX)$ are also compensated by
the Green-Schwarz mechanism. 
All the anomaly coefficients which are quadratic in $X$ vanish. 
Obviously, we have $(XX SU(2))=(XX SU(3))=0$ and $(XX Y_F)=0$ 
(as usual, the vector-like Higgs do not contribute).
Again, the fact that $U(1)_X$ does not lie into $SU(5)$
guarantees $(XXY)=0$. 

There are also gravitational anomalies. 
In particular, $C_{grav} \equiv (XTT) \neq 0$, 
where $T$ is the energy-momentum tensor, is also compensated 
by the Green-Schwarz mechanism.
All the remaining possible anomaly coefficients involving $T$ vanish.

\section{Conclusions}

Neutrino physics is a privileged sector for the  study of new physics beyond
the Standard Model. 
The accumulated neutrino oscillation data over the last few years
are precise enough to give serious hints on the structure
of a possible unifying theory.
They not only indicate the existence of neutrinos masses, 
the see-saw mechanism also provides a unique opportunity to glimpse 
some mysteries at scales as high as the unification scale.

In the spirit of a hierarchical unifying structure, we have found
that a simple neutrino mass matrix is suggested by the present data
on neutrino oscillations. It can accommodate two large and one small
mixing angles without strong fine-tuning.

In the language of an effective theory with Abelian family symmetries,
the lepton doublets charges derived from this matrix,
when combined with the hierarchy and mixing structure observed in
the quark sector, give a strong indication that the family symmetries lie
beyond $SU(5)$. 
With the explicit construction of such a model, we see that this simple
picture is indeed in a remarkable agreement with all the present data 
on quarks and leptons masses and mixings. Moreover, the model is
by itself theoretically consistent and economic. It exhibits
anomaly cancellation, and the correct energy scales for the string
cut-off and for leptogenesis are obtained.

\section*{Acknowledgments}

We are thankful to Stuart Wick for helpful discussions throughout this work, which 
is supported by the United States Department of Energy
under grant DE-FG02-97ER41029.

\section*{{\large Appendix: Mixing matrix and eigenvalues of 
$M_\nu^{(2)}$ for $D_0 \sim O(\lambda^2)$ and $D_1 \sim O(\lambda^2)$}}

The formulas given below assume that the expansion in $\lambda$ is
valid. For simplicity, we also neglect CP violating phases, so
that all the matrix elements are real, and the mixing matrix is
orthogonal.
Then, for $D_0 \sim O(\lambda^2)$ and $D_1 \sim O(\lambda^2)$,
$M_\nu^{(2)}$ can be parameterized as
\be
M_\nu^{(2)} = m_0 \left(
\ba{ccc}
a \lambda^2 & d \lambda & f \lambda \\
d \lambda & b & e \\
f \lambda  & e & c \\
\ea
\right)
\ee
with
\bean
b &=& \gamma \sin^2 \theta + b_1 \lambda + b_2 \lambda^2 \\
c &=& \gamma \cos^2 \theta + c_1 \lambda + c_2 \lambda^2 \\
e &=& \gamma \sin \theta \cos \theta 
+ e_1 \lambda + e_2 \lambda^2 \\
d &=& \beta \sin \theta + d_1 \lambda + d_2 \lambda^2 \\
f &=& \beta \cos \theta + f_1 \lambda + f_2 \lambda^2 \\
\eean
where all the coefficients $\theta$, $\gamma$, $\beta$, $b_1$,...
are supposed to be of order $O(1)$.

The condition $D_1 \equiv \delta _1 \lambda^2$ gives the relation
\be
\cos^2 \theta \, b_1 +\sin^2 \theta \, c_1
-2 \sin \theta \cos \theta \, e_1 = 0 
\ee
and 
\be
\delta _1 = \delta'_1 + \gamma \delta'_2
\ee
with
\bean
\delta'_1 &=& b_1 c_1 - e_1^2 \\
\delta'_2 &=& \cos^2 \theta \, b_2 +\sin^2 \theta \, c_2
-2 \sin \theta \cos \theta \, e_2 \\
\eean
It can then be verified that $D_0 \equiv \delta _0 \lambda^2$ with
\be
\delta _0 = a \delta _1 - \gamma \kappa^2 - 2 \beta \kappa \sigma 
-\beta^2 \delta'_2
\ee
and
\bean
\kappa &=& d_1 \cos \theta - f_1 \sin \theta \\
\sigma &=& e_1 - b_1 \cot \theta \\
\eean

The eigenvalues of $M_\nu^{(2)}$ are given by
\be
\mu _{0,1} \simeq m_0 \; k_{0,1} \lambda^2
\hspace{1cm}
\mu _2 \simeq m_0 \; \gamma 
\ee
with
\be
k_{0,1} = \frac{(E+\delta _1) \pm 
\sqrt{(E+ \delta _1)^2-4 \gamma \delta _0}}{2 \gamma}
\ee
and $E = a \gamma -\beta^2$.

The mixing matrix is given by
\be
U \simeq \left(
\ba{ccc}
\cos \phi & \sin \phi & \frac{\beta}{\gamma} \lambda \\
-\sin \phi \cos \theta & \cos \phi \cos \theta & \sin \theta \\
\sin \phi \sin \theta & -\cos \phi \sin \theta & \cos \theta \\
\ea
\right)
\ee
with
\be
\tan \phi = \frac{a \gamma - k_0 \gamma - \beta^2}
{\gamma \kappa + \beta \sigma} = 
-\frac{\gamma \kappa + \beta \sigma}
{a \gamma - k_1 \gamma - \beta^2}
\ee
Consistency for the value of $\tan \phi$ requires that
\be
(a \gamma - k_0 \gamma - \beta^2)(a \gamma - k_1 \gamma - \beta^2)
+ (\gamma \kappa + \beta \sigma)^2 = 0
\ee
which indeed can be verified using the fact that $\sigma^2 = -\delta'_1$.

\end{document}